\documentclass[a4paper,12pt,times new roman]{article}
\usepackage[top=1in, bottom=1in, left=24mm, right=24mm]{geometry}
\usepackage{graphicx}
\usepackage{epsfig}
\usepackage{lineno}
\usepackage{setspace}

\begin{document}
\title{\bf{Space charge and collective oscillation of ion cloud in a linear Paul trap}} \vspace{3pt}
\author{\bf{P. Mandal\footnote{Present Address: Department of Physics, St. Paul's
Cathedral Mission College, 33/1 Raja Rammohan Roy Sarani, Kolkata
700 009, India. Email: \emph{pintuphys@gmail.com}}, S. Das$^{\dag}$,
D. De Munshi$^{\dag}$, T. Dutta$^{\dag}$ and M.
Mukherjee}\footnote{Present Address: Centre for Quantum
Technologies, National University of Singapore,
Singapore - 117543} \\
\\
Raman Center for Atomic, Molecular and Optical Sciences \\
Indian Association for the Cultivation of Science \\
2A \& 2B Raja S. C. Mullick Road, Kolkata 700 032 \\
\\
} \maketitle
\begin{center}
\textbf{Abstract} \\
\end{center}


The presence of charged particles in an ion trap modifies the
harmonic trapping potential in which they are trapped, leading to
observed shifts in secular frequency as well as appearance of
collective oscillation. In a linear trap geometry, both of these
effects have been observed under different trapping conditions using
narrow non-linear resonance and external excitation. The
observations have been modeled with minimal fitting parameter
showing good agreement with results obtained. The space charge in
our experiment plays an important role in terms of criticality of
the onset of collective oscillation.

\vspace{1cm}


\section{Introduction}\label{section1}
Ion trap finds numerous applications in physics,
chemistry, biology and engineering. In some cases a few ions are
used to carry out high precision
experiments~\cite{Lin09,Berkeland98,Raizen92,Roos06} while in other
cases high density of ions are required to store in the
trap~\cite{Church69}. However, in both of these experiments Coulomb
repulsion among the ions needs to be controlled or understood well
enough to either estimate systematic or reduce the effect. In spite
of numerous studies on space charge and its effect on the trapped
ions, some of the observed phenomena are yet to be understood from
ab-initio theory. This includes collective oscillation that has been
observed both in Penning~\cite{Wineland75} as well as in Paul trap
of hyperbolic structure~\cite{Alheit97}. In this article we report
on space charge in a linear Paul trap and its effect on the ion
dynamics in two different manifestations. First, it modifies the
individual ion oscillation frequency due to the modification of the
trapping potential; second, under certain experimental condition of
high charge density the ions behave as a single collective body
showing motional frequency independent of the number of ions. We put
forward a model to account for the space charge effect and show that
it explains well the observed shifts of the individual ion
oscillation frequency. In respect to the observed collective
oscillation of the radial motion, our result shows similar behaviour
as that was observed earlier in the collective oscillation of the
axial mode of ref.~\cite{Alheit97}. In addition to the excitation
energy criticality, we observe criticality of the collective
oscillation on ion number which needs to be explained theoretically.

In the theoretical section, the relevant ion trap theory will be
briefly reviewed followed by the experimental setup. The ``result
and analysis" section contains the experimental data and analysis
with the space charge model. The observation of collective
oscillation of the radial modes in a linear Paul trap and its
criticality with respect to the ion number, have been shown in this
section.

\section{Theory}

A linear Paul trap uses a radio frequency (rf) potential
($V_{0}\cos\Omega t$) in addition to a dc potential ($U$) for
providing dynamical trapping of charged particles in the radial
plane and a dc potential for axial confinement. The radial potential
in terms of the radial coordinate ($r,\phi$), can be expressed
by~\cite{Dawsonbook}
\begin{equation}\label{eqn1}
\Phi(r,\phi,t) = \sum_{k}a_{k}\left(\frac{r}{r_{0}}\right)^{k}\cos
k\phi(U-V_{0}\cos\Omega t),
\end{equation}
where $r_{0}$ is the distance of an electrode from the trap center
and $a_{k}$ is the weight factor for $k$th order multipole. The
equation of motion of a trapped ion corresponding to the quadrupole
part ($k=2$) of the potential in eqn.~\ref{eqn1} is given by the
Matheiu's differential equation as,
\begin{equation}\label{eqn2}
\frac{d^{2}u}{d\zeta^{2}} + (a_{u}-2q_{u}\cos2\zeta)u = 0,
\end{equation}
with $u=x,y$, where
\begin{eqnarray}\label{eqn3}
a_{x}&=&-a_{y}=\frac{4eU}{mr_{0}^{2}\Omega^{2}}, \nonumber \\
q_{x}&=&-q_{y}=\frac{2eV_{0}}{mr_{0}^{2}\Omega^{2}},
\end{eqnarray}
and $\zeta=\Omega t/2$. Solution of eqn.~\ref{eqn2} shows that the
ion oscillates with two different frequencies~\cite{Major05}: one
equal to the frequency of the applied rf and is called micromotion
while the other with a frequency $\omega_{0u}=\beta_{u}\Omega/2$ and
is called the secular motion. Here
$\beta_{u}\approx\sqrt{a_{u}+q_{u}^{2}/2}$ within the adiabatic
approximation \emph{i.e.} for small $a_{u}$ and
$q_{u}$~\cite{Herfurth01}.

The trapped ion performs in-phase simple harmonic oscillation due to
the quadrupole potential but gains energy from higher order
multipoles that enhances their motional amplitude. The ion motion
gets resonantly excited when the frequency of oscillation at a given
operating parameter ($a_{u}$, $q_{u}$) satisfies the following
condition~\cite{Busch61,Dawson69,Wang93}
\begin{equation}\label{eqn4}
n_{x}\omega_{0x}+n_{y}\omega_{0y}=\Omega,
\end{equation}
where $n_{x},n_{y}=0, 1, 2, 3...$ and $n_{x}+n_{y}=k$ and $>2$. If
one of the trap parameters is varied, it will modify the $\beta$
value and hence the secular frequency. Thus a nonlinear resonance
(NLR) appears at a definite value subjected to the condition defined
by eqn.~\ref{eqn4} and results in narrow instabilities within the
broad stability diagram~\cite{March05}. As the trap operating
parameters get modified due to numerous factors like, the coupling
between the axial and radial potentials~\cite{Drakoudis06} and space
charge due to the trapped ions~\cite{Alheit96}, the NLR over
$q-$parameter-space shifts. The amplitude of such resonances is
determined by the strength of the corresponding multipole which
strongly depends on the geometry of the setup. The geometry
dependence of higher order multipoles has been studied in detail in
mass filters~\cite{Reuben96} and linear traps with cylindrical
rods~\cite{Pedregosa10}. Here we discuss the effect of space charge
on NLR of an ion cloud in a linear Paul trap of particular geometry
described in Sect.~\ref{sec:1}.

A simple model has been employed to explain the observed shift in
the nonlinear resonance as reported in this article as well as other
experiment reported elsewhere~\cite{Alheit96}. The effective
potential in the vicinity of a trapped ion is the space charge
potential developed by all other trapped ions, in addition to the
applied potential. Since the space charge potential is developed by
the trapped ions themselves, the effective potential experienced by
an ion grows with the density of trapped ions ($\rho$). In this
model, it is assumed that the space charge potential is proportional
to the density of trapped ions and a quadratic function of the
position coordinate as consistent with Poisson's equation. Thus the
space charge potential in the radial plane can be represented by
\begin{equation}\label{eqn6}
\Phi_{s}=-\kappa\rho(x^{2}+y^{2}),
\end{equation}
where $\kappa$ is a constant. Negative sign in eqn.~\ref{eqn6} is
justified by the positive type charge in space. Incorporating the
space charge potential in the applied trapping
potential~(eqn.~\ref{eqn1}), the equation of motion of an ion in the
radial plane can be written following eqn.~\ref{eqn2} as
\begin{equation}\label{eqn7}
\frac{d^{2}u}{d\zeta^{2}} + (a'_{u}-2q_{u}\cos2\zeta)u = 0,
\end{equation}
where
\begin{equation}\label{eqn8}
a'_{u}=a_{u}-\frac{8\kappa\rho e}{m\Omega^{2}}.
\end{equation}

The definition of the $\beta$ parameter, within adiabatic
approximation, is modified by $a'_{u}$ and can be redefined by
\begin{equation}\label{eqn9}
\beta'_{u}=\sqrt{a'_{u}+\frac{q_{u}^{2}}{2}}.
\end{equation}

Thus the secular frequency of the trapped ion
decreases~\cite{Enders92} as $\beta$ parameter is effectively
decreased due to the space charge effect. In order to match the
secular frequency in presence of the space charge, the NLR shifts to
higher value of $q$. If it shifts to a new position $q'$ in presence
of the space charge, it is given by
\begin{equation}\label{eqn10}
q'=\sqrt{2(a-a')+q_{0}^{2}},
\end{equation}
where $q'=q_{0}$ is the NLR center in absence of the space charge
effect. From eqns.~\ref{eqn8} and \ref{eqn10}
\begin{equation}\label{eqn11}
q'=\sqrt{\frac{16\kappa\rho e}{m\Omega^{2}}+q_{0}^{2}}.
\end{equation}

\begin{figure*}
\hspace{-1.2cm} \vspace{-1cm}
 \resizebox{1.2\textwidth}{!}{\rotatebox{270}{
  \includegraphics{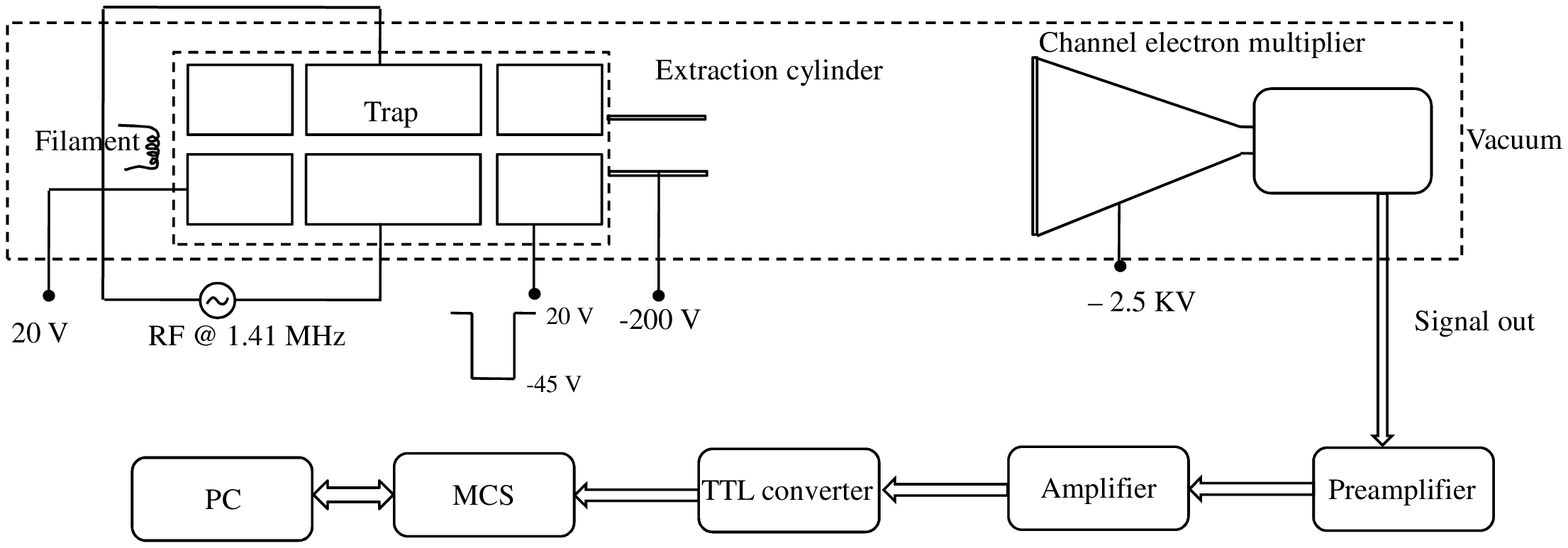}
 }}
 \vspace*{-6cm}
\caption{Schematic of the experimental setup. The filament, trap,
channel electron multiplier with other ion optics are housed inside
a vacuum chamber at pressure $10^{-7}$~mbar. The electronics outside
the vacuum are used for generating the ToF signal.} \label{fig1}
\end{figure*}

\section{Experiment}\label{sec:1}

The experimental setup consisting of a linear Paul trap, an
ionization setup, extraction and detection setup is shown
schematically in figure~\ref{fig1}. The linear trap is assembled
with four three-segmented cylindrical rods (see for example,
ref.~\cite{Drewsen00}), each of diameter $10$~mm with a separation
of $8$~mm between two opposite electrodes~($2r_{0}$). The middle
electrodes are taken $25$~mm in length. Molecular nitrogen ions
(N$_{2}^{+}$) are created from the background gas at pressure
$10^{-7}$~mbar by electron impact ionization. The ions are
dynamically trapped for few hundreds milliseconds to seconds before
they are extracted by lowering the axial potential in one direction.
Typical time-of-flight (ToF) obtained via a channel electron
multiplier (CEM) and other electronics shown in schematic
(figure~\ref{fig1}), is $8$~$\mu$s with full-width at half-maxima of
$12$~$\mu$s and this is well within the saturation limit for a
typical CEM detector. The ions of different charge-to-mass ratio can
be identified from the ToF~\cite{Alheit96}.

The trap is operated at a frequency of $\Omega/2\pi=1.415$~MHz and
the middle electrodes are kept at dc ground ($U=0$, $a_{u}=0$). The
operating parameter $q$ is varied by changing the rf amplitude while
keeping the ion creation time (T$_{c}$), ion storage time and all
other parameters constant. The experiment at a given $q$ is repeated
$100$ times and the total number of trapped ions is counted. As
expected, narrow NLRs appear within the stability diagram due to the
presence of higher order multipoles of the trap potential. The
observed resonances in our setup correspond to the $6$th, $7$th and
$8$th order multipoles. However, as the sixth order is the strongest
resonance in our setup, it has been used for further experiments
described in this article.

\begin{figure}
\centering 
\resizebox{0.55\textwidth}{!}{\rotatebox{270}{%
  \includegraphics{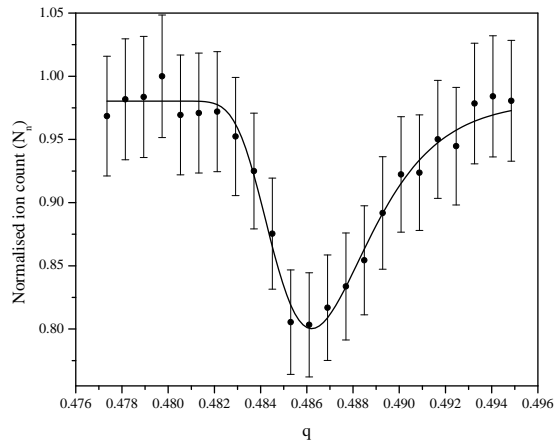}
}} \caption{The normalized ion count as a function of $q$ that shows
  an NLR corresponding to sixth order multipole in our setup. The solid line is
  a fit to the data as explained in the text.} \label{fig2}
\end{figure}

The $q$ parameter is varied in steps of $0.0004$ about the sixth
order NLR keeping all other parameters unchanged. Experimentally
obtained ion numbers are normalized with respect to the maximum ion
number ($N_{0}$) during a particular resonance experiment. The
normalized ion count ($N_{n}$) with the associated uncertainty has
been plotted as a function of $q$ as shown in figure~\ref{fig2},
which corresponds to a maximum number of trapped ions around $170$.
This resonance (figure~\ref{fig2}) is fitted with a Gumbel function
as defined by
\begin{equation}\label{eqn5}
N_{n}=N'_{0}+A\exp\left[-\exp(-\eta)-\eta+1\right],
\end{equation}
where $\eta=(q-q')/\sigma$. Here $q'$ is the NLR center and $\sigma$
is a scale parameter. This Gumble fit is justified by the extreme
nature of distribution of the ions which are excited and rejected
due to the NLR for a given trap acceptance. As can be seen from
figure~\ref{fig2}, the fitting curve (shown by solid line) lies
within the experimental error bars with the adjacent $R-$square
value of $0.98$. The position of the resonance as obtained from this
fit is $q'=0.4862$($4$) and $\sigma=0.0021$($1$). The same fit
routine has been used for consistency to obtain the resonance center
in the subsequent analysis.

\section{Results and analysis}

\subsection{Shift in the resonance center}\label{sec:2}

\begin{figure}
\centering 
\resizebox{0.55\textwidth}{!}{\rotatebox{270}{%
  \includegraphics{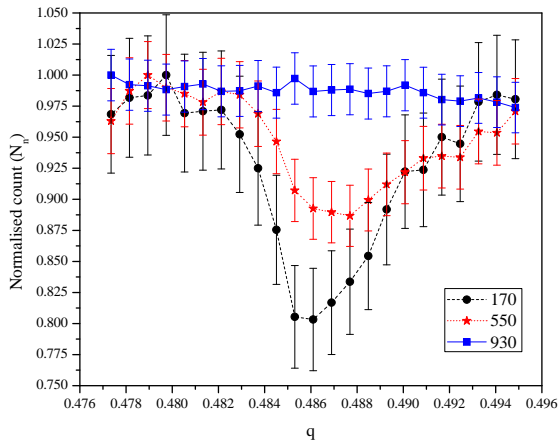}
}} \caption{Normalized ion count
  as a function of the trap operating parameter $q$ ($a=0$) for different
  number of trapped ions (shown inset). Only three representative resonances
  out of six sets are shown here for clarity.} \label{fig3}
\end{figure}

\begin{figure}
\centering 
\resizebox{0.55\textwidth}{!}{%
  \includegraphics{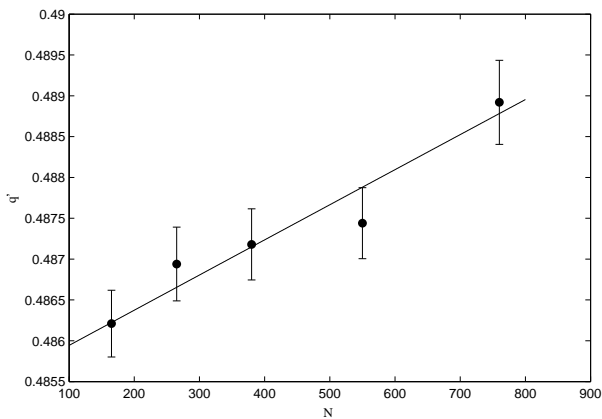}
} \caption{Shift in the NLR center ($q'$) with trapped ion number
($N$). Solid line is a fit to the data points with the model
function as described in the text.}
\label{fig4}       
\end{figure}

The number of trapped ions has been varied by increasing the
creation time T$_{c}$, and the same NLR as described in
Sect.~\ref{sec:1} has been studied for each set (\emph{i.e.} at a
given T$_{c}$). The NLR corresponding to different number of trapped
ions are obtained and only three of them have been presented in
figure~\ref{fig3}. The figure shows that the strength of the NLR
becomes weaker and broader as more ions are loaded into the trap.
When the trap is loaded with its full capacity (corresponds to
maximum trapped ion number of around $930$ shown in
figure~\ref{fig3} and known from the loading capacity of the trap),
the signature of the NLR disappears. Each of the NLRs corresponding
to different ion numbers has been fitted to the function described
in eqn.~\ref{eqn5} in order to determine the resonant $q$ value
($q'$). The $q'$ has been plotted as a function of the number of
trapped ions ($N$) in figure~\ref{fig4} which shows that it shifts
to higher values when $N$ is increased.

Assuming the effective size of the ion cloud ($\tau_{0}$) unchanged
as more ions are loaded into the trap, the charge density scales
linearly with the number of trapped ions ($N$). Thus
eqn.~\ref{eqn11} can be represented in terms of $N$ as
\begin{equation}\label{eqn12}
q'=\sqrt{bN+c},
\end{equation}
with parameters $b$ and $c$ defined as $b=16\kappa
e^{2}/m\tau_{0}\Omega^{2}$ and $c=q_{0}^{2}$ which are constant for
a given species and trapping condition.

The shift of the NLR center with the trapped ion number as presented
in figure~\ref{fig4}, has been fitted with the model function
described in eqn.~\ref{eqn12}. However, due to the small trap volume
the values of $N$ are not widely spread and hence the fit resembles
a straight line. The adjacent $R-$square value of the fit is $0.9$
and it yields $b = 4.2(2)\times10^{-6}$ and $c = 0.2357(5)$. The fit
routine yields the size of the ion cloud
$\tau_{0}\approx1.5$~mm$^{3}$ (with $\kappa=1/2\epsilon_{0}$) and an
unperturbed value of $q$ as $q_{0}=0.4855(5)$, the later agrees well
with the resonant $q-$value for low ion number where the space
charge does not contribute significantly.

\begin{figure}
\centering 
\resizebox{0.55\textwidth}{!}{\rotatebox{270}{
  \includegraphics{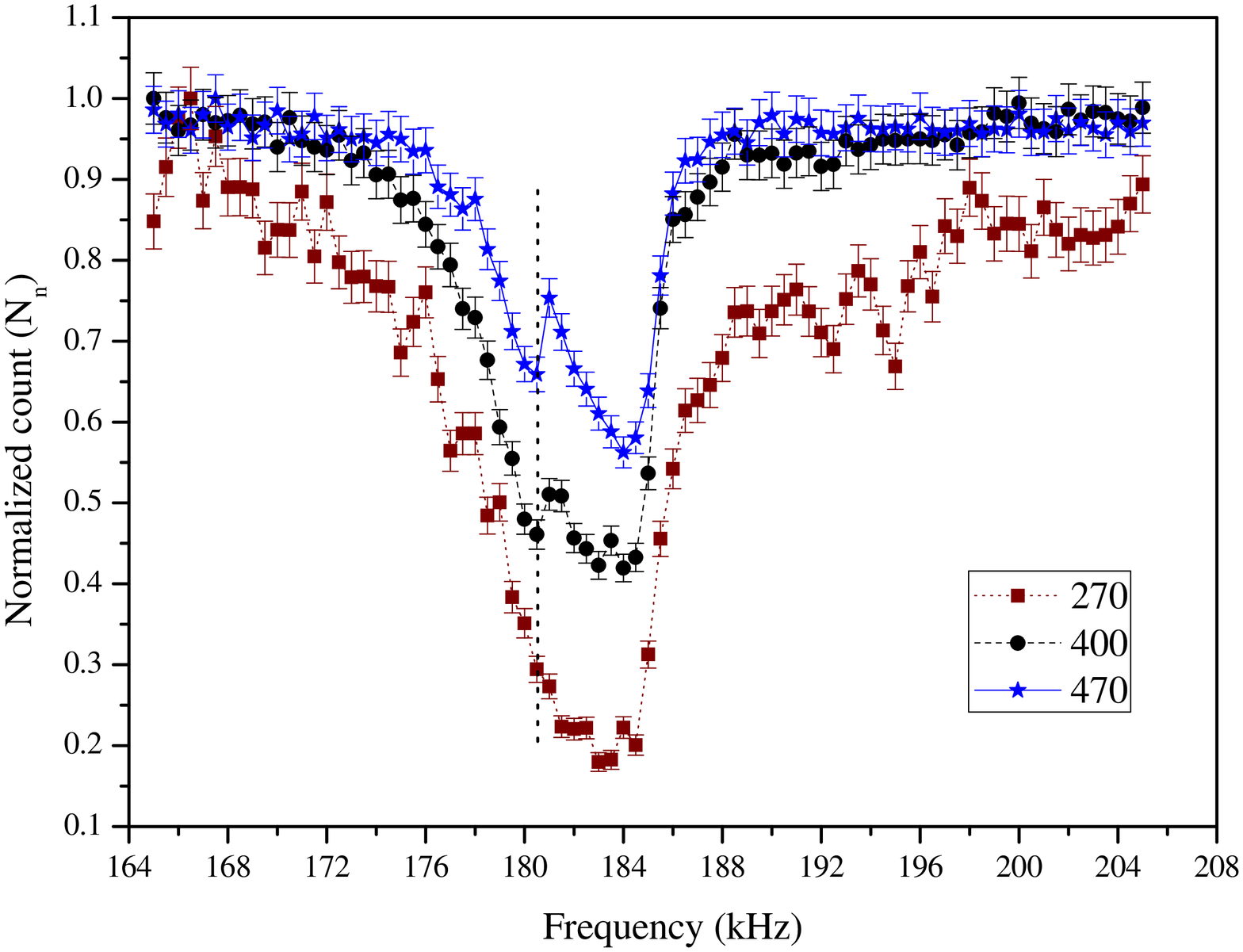}
 }}
\caption{Normalized ion number
  as a function of the frequency of the electric dipole excitation
  field for different number of trapped ions (mentioned inset). The
  second dip at $180.5$~kHz and marked by the dotted line in the resonance
  profiles for estimated trapped ions of $400$ and
  $470$, corresponds to the collective oscillation.} \label{fig5}
\end{figure}

\subsection{Amplitude of resonance}\label{sec:4}

One of the important observations in our experiment is the
suppression of NLR~(figure~\ref{fig3}) when the trap is loaded
almost to its full capacity (Sect.~\ref{sec:2}). This phenomenon can
be qualitatively explained from the emergence of collective
oscillation~\cite{Alheit97,Jungmann87,Vedel90} observed in our trap
for the radial mode of motion. As the coupling between the ions by
inter-ionic Coulomb interaction grows due to large ion number, the
cloud behaves as an effective single ion in the trap. The effective
charge of the collectively oscillating cloud being higher, requires
higher excitation energy. Moreover, isotropy of the cloud under
collective oscillation does not contribute to higher order
nonlinearity unlike uncoupled ion distribution. Thus there are less
reasons for NLRs. The strength of the higher order multipoles are
relatively weak in our linear trap as observed from
figure~\ref{fig3}, and not sufficient to excite the collective
oscillation of the ion cloud resulting in disappearance of the
resonance at larger number of ions.

\subsection{Collective oscillation}\label{sec:5}

Collective oscillation of the trapped ions in the radial plane
appears prominent (figure~\ref{fig5}) when they are excited by an
externally applied electric dipole field, the strength of which can
be controlled. Figure~\ref{fig5} shows that a second dip at
$180.5$~kHz appears for estimated trapped ion number $>270$ due an
electric dipole excitation amplitude of only $50$~mV. This
corresponds to the excitation of the collective oscillation.
Individual ions in the normal cloud would see the space charge
formed by their neighbours which leads to a shift in the resonance
center as discussed before. This is, however, not the case for
collective oscillation as it is independent of the ion number above
a threshold. Figure~\ref{fig5} shows that the collective oscillation
gets excited and the resonance becomes more prominent with
increasing ion number whereas the excitation of individual motion
becomes weaker as more ions oscillate collectively.

\section{Conclusion}\label{Sec:6}

The effect of space charge on the NLR of an ion cloud in a linear
Paul trap leading to frequency shift and emergence of collective
oscillation, has been reported in this article. It is explained with
the help of a model which is analytically consistent with earlier
report where the trapped ion number has been widely
varied~\cite{Alheit96}. Though the actual spatial distribution of
trapped ions is Gaussian~\cite{Knight78,Schaaf81}, the consideration
of a simple uniform distribution as is done here, is equally
efficient to explain small variation of the NLR line center. It
shows that the space charge potential in the vicinity of a single
ion within the trap can be considered as an effective dc potential.
Such a model has been considered earlier as well to estimate the
number of trapped ions~\cite{Schuessler69,Fulford80}. The effective
size of the ion cloud as obtained from the experiment is
$1.5$~mm$^{3}$; or in other words, the cloud is spread over $10\%$
of the radial and axial dimension of the trap which is the typical
size of `hot' ion cloud as consistent with ref.~\cite{Cutler85}.
Again considering the maximum number of trapped ions around $1000$
as obtained from the study on the loading capacity of the trap, the
maximum density of the trapped ions has been estimated of the order
of $10^{3}$ per mm$^{3}$.

Contrary to earlier report~\cite{Alheit96}, vanishing NLR as a
function of space charge has been observed. Apart from the
possibility of collective oscillation taking over the individual
oscillations, the contradiction may stem from stronger
higher-order-multipoles in the hyperbolic trap~\cite{Alheit96} as
compared to that in our linear trap. The experiment~\cite{Alheit96}
has been performed with a fourth order NLR, the strongest one in the
hyperbolic trap while with the sixth order in our linear trap. It is
also not clear whether the hyperbolic trap which has larger volume
allowing more ions to be trapped, is loaded with its maximum
capacity as done in our setup.

The adiabatic approximation has been applied here to explain the
observed shift of the sixth order resonance around $q=0.4855$. It
could be interesting to verify the model for lower values of $q$ as
well. However, the strength of higher order resonance that appears
below $q=0.4$ is found very weak in our setup and hence the required
resolution of the resonant $q$ values as a function of the number of
ions has not been achieved. The signature of collective oscillation
has not been observed via NLR but it is quite prominent in the
electric dipole excitation experiment leading to the following two
reasons behind this non-observation. (1) The width of the NLR is
broader than the narrow collective oscillation resonance and hence
the later remains hidden within the former. (2) The strength of
higher order multipole (here, the sixth order) is not strong enough
to excite collective oscillation of a large ion cloud. Both of these
possibilities need further experimental and theoretical
investigations.

\section{Acknowledgements}

P.~Mandal is thankful to G.~Werth, University of Mainz, Germany for
helpful discussion on the interpretation of the observations of the
experiment.

\end{document}